\documentclass[5p]{elsarticle}

\usepackage{graphics,graphicx}
\usepackage{amsmath,amsfonts,amssymb}
\usepackage{algorithm}
\usepackage{algpseudocode}
\usepackage{color}
\usepackage{lineno,hyperref}
\newcommand{\thB}{\mbox{\boldmath$\theta$}}

\newcommand{\bb}{\mathbf{b}}

\journal{Atmospheric Environment}

\begin{document}

\begin{frontmatter}

\title{Localisation of a source of biochemical agent dispersion using binary measurements} %\tnoteref{t1}}
%\tnotetext[t1]{\copyright Commonwealth of Australia}



\author[dsto]{Branko Ristic\fnref{fnBranko}}

\author[dsto]{Ajith Gunatilaka\corref{mycorrespondingauthor}}
\ead{ajith.gunatilaka@dsto.defence.gov.au}

\author[dsto]{Ralph Gailis\fnref{fnRalph}}

\cortext[mycorrespondingauthor]{Corresponding author}

\fntext[fnBranko]{Now with RMIT University}

\fntext[fnRalph]{Now with NSID, DST Group}
\address[dsto]{Land Division, DST Group, Melbourne, Australia}

\begin{abstract}
Using the measurements collected at a number of known locations by a moving binary sensor,
characterised by an unknown threshold, the problem is to estimate
the parameters of a biochemical source, continuously releasing material into the
atmosphere.  The solution is formulated in the Bayesian framework
using a dispersion model of Poisson distributed particle encounters
in a turbulent flow.  The method is implemented using the importance
sampling technique and successfully validated with three
experimental datasets under different wind conditions.
\end{abstract}


\end{frontmatter}

%\linenumbers

\section{Introduction}
\label{sec:intro}

Localisation of a source of biological or chemical agent dispersing
in the atmosphere is an important problem for national security
and environmental monitoring applications \cite{kendal}. Wind, as
the dominant transport mechanism in the atmosphere, %exhibits%
can generate strong turbulent motion, causing the released agent to
disperse as a plume whose spread increases with the downwind
distance \cite{arya_98}. Assuming a constant release of the
contaminant, the problem involves estimation of source parameters:
its location and intensity (release-rate). Two types of measurements
are generally at disposal for source localisation: (i) the
concentration measurements at spatially distributed sensor
locations; (ii) the average wind speed and wind direction (typically
available from a nearby meteorological station).

Many references are available on the topic of biochemical source
localisation, assuming un-quantised (analog) concentration
measurements. Standard solutions are based on optimisation
techniques, such as the nonlinear least squares \cite{matthes_05} or
simulated annealing \cite{thomson_07}. These methods are unreliable
due to local minima or poor convergence; in addition, they provide
only point estimates, without uncertainty intervals. The preferred
alternative is the use of Bayesian techniques; they result in the
posterior probability density function (PDF) of the source parameter
vector, thereby providing an uncertainty measure to any point
estimate derived from it. Most Bayesian methods for source
estimation are based on Markov chain Monte Carlo (MCMC) technique,
assuming either Gaussian or log-Gausiian likelihood function of
measurements \cite{keats_07,humphries_12,ortner,senocak_08}.
Recently, a likelihood-free Bayesian method for source localisation
was proposed in \cite{ristic_15}.

Binary sensor networks have become widespread in environmental monitoring applications
because binary sensors generate as little as one bit of information.
Such binary sensors allow inexpensive sensing with minimal communication
requirements \cite{aslam_03}. In the context of binary sensor
networks, an excellent overview of non-Bayesian chemical source
localisation techniques is presented in \cite{chen2008greedy}. Best
achievable accuracy of source localisation using binary sensors has
been discussed in \cite{Ristic2014}.

Prior work in using binary sensor data for biochemical source localisation assumes that the detection threshold
of the sensor is known. It is a reasonable assumption for a commercial sensor whose sensitivity is specified (for example, in  parts per
million by volume (ppm$_v$) or grams per cubic meter) by the manufacturer and when the sensor is well calibrated. However, we consider at least two scenarios where the detection threshold of a binary sensor may not be accurately known. The first scenario is when a sensor's detection threshold goes off calibration due to environmental conditions such as temperature or humidity or ageing of the sensor. The second scenario is where the sensor is a human rather than a device. For example, imagine a person smelling a strong odour such as due to a gas leak or a decomposing animal carcass. When the person moves around, the smell will be detected in some locations but not in others, producing a binary measurement sequence without knowing the exact value of the threshold in ppm or g/m$^3$. In this paper, we develop a Bayesian algorithm that carries out source parameter estimation based on such
{\em binary concentration measurements} where the sensor threshold is unknown. A Monte Carlo technique, importance sampling, is applied to calculate the posterior PDF approximately. The method is successfully
validated using three experimental datasets obtained under different wind conditions.

\section{Models}
\label{s:2}

\subsection{Dispersion model}
To solve the source localisation problem described above, we propose a solution  formulated in the Bayesian framework which relies on
two mathematical models: the atmospheric dispersion model and the
concentration measurement model. A dispersion model mathematically describes the physical processes that govern the
atmospheric dispersion of the released agent within the plume. The
primary purpose of a dispersion model is to calculate the mean
concentration of emitted material at a given sensor location. A
plethora of dispersion models are in use today
\cite{holmes_morawska_06} to account for specific weather
conditions, terrain, source height, etc. In this paper, we adopt a
two-dimensional dispersion model of ``particle encounters'' in a
turbulent flow, described in \cite{vergassola_07}. During a certain sensing
period, each sensor  experiences a Poisson distributed number of
``encounters'' with released particles. The binary nature of
measurements indicates that a sensor reading of binary ``1'' or a ``positive detection''  corresponds to
the number of such encounters exceeding a particular threshold.

 If a binary sensor with a particular threshold makes
positive detections (binary ``1'') at some locations and zero detections (binary ``0'') at other locations due to a source of 
a certain release rate, the measurements at these locations will be the same even if both the source release rate and the sensor detection
threshold were scaled up or down together by the same amount; it is  the ratio between the source release rate and the sensor threshold that determines 
which sensor locations will have positive or zero readings. Therefore, when we estimate the source parameters using binary data from a sensor whose
detection threshold is unknown, it is not possible to estimate the absolute value of the source release rate; only the release rate normalised by the assumed sensor threshold can be estimated. Nevertheless, the source location, which is actually the parameter of main interest, can be estimated.  Without loss of generality, in our 
experiments, we assumed the sensor to output binary ``1'' if it encounters at least one particle during a sensing period and output a binary  ``0'' otherwise. 

Let us assume that the biochemical source is located at  $(x_0, y_0)$, with a normalised release rate of $Q_0$.  The particles released from the source propagate with the isotropic diffusivity $D$, but can  also be advected by wind. We assume the released particles to have an average lifetime of  $\tau$. While the wind speed is typically available from meteorological data from a nearby measuring station, we use this speed as the prior guess for a Bayesian estimate of the true, effective wind speed affecting the advection of particles. Accordingly, let us  assume that the mean wind speed is $V$ and  the mean wind  direction coincides with the direction of the $x$ axis. We denote the PDF of the wind speed by $\pi(V)$.  A spherical sensor of
small size $a$ at a location with coordinates $(x,y)$, non-coincidental with the source location $(x_0,y_0)$, will experience a series
of encounters with the released particles.

Denoting the parameter vector we wish to estimate, consisting of the source coordinates  ($x_0$, $y_0$), normalised source release rate $Q_0$, and the wind speed $V$, by  $\thB = [x_0\;y_0\;
Q_0\; V]^\intercal$, the rate of particle encounters by the sensor at the  $i$th location (where $i=1,\dots,M$) with coordinates $(x_i,y_i)$ can be modelled
as \cite{vergassola_07}:
\begin{equation}
R(x_i,y_i|\thB) =
\frac{Q_0}{\ln\left(\frac{\lambda}{a}\right)}\,\exp\left[{\frac{(x_0-x_i)V}{2D}}\right]\cdot
K_0\left(\frac{d_i(\thB)}{\lambda}\right)  \label{e:disp}
\end{equation}
where $D$, $\tau$ and $a$ are known environmental and sensor parameters,
\begin{equation}
d_i(\thB) = \sqrt{(x_i-x_0)^2+(y_i-y_0)^2}
\end{equation}
is the distance from the source to  $i$th sensor location, $K_0$ is the modified
Bessel function of order zero, and
\begin{equation}
\lambda = \sqrt{\frac{D\tau}{1+\frac{V^2\tau}{4D}}}.
\end{equation}

\subsection{Measurement model}
The stochastic process of sensor encounters with released particles
is modelled by a Poisson distribution. The probability that 
sensor at location $(x_i,y_i)$ encounters
$z\in\mathbb{Z}^+\cup\{0\}$ particles ($z$ is a non-negative
integer) during a time interval $t_0$ is then:
\begin{equation}
\mathcal{P}(z; \mu_i) = \frac{(\mu_i)^{z}}{z!}e^{-\mu_i}
\label{e:likf}
\end{equation}
where $\mu_i =  t_0\cdot R(x_i,y_i|\thB) $ is the mean concentration
at $(x_i,y_i)$. Equation (\ref{e:likf}) represents the full
specification of the likelihood function of parameter vector $\thB$,
given the sensor encounters $z$ counts at the  $i$th position.

However, because the actual sensor is binary, the measurement model is
\begin{equation}
b_i = \begin{cases} 1, & \text{if } z =1,2,3,\dots\\
0, & \text{if } z = 0.
\end{cases}
\end{equation}
Note that $b_i$ is a Bernoulli random variable with the parameter
\begin{eqnarray} q_i(\thB) & = & \text{Pr}\{b_i=1\}\\
   & = & \sum_{z=1}^{\infty} \mathcal{P}(z;\mu_i) \\
   & = & 1 -  \mathcal{P}(0;\mu_i) \\
   & = & 1 - e^{-\mu_i}.
   \end{eqnarray}
The likelihood function for the sensor when it is at the $i$th location is then:
\begin{equation}
p(b_i|\thB) = [q_i(\thB)]^{b_i} \, [1-q_i(\thB)]^{1-b_i}.
\end{equation}
Assuming sensor measurements are conditionally independent, the likelihood
function of the parameter vector $\thB$,  given the binary
measurement vector $\bb = [b_1,\dots,b_M]^\intercal$, is a product:
\begin{equation}
p(\bb|\thB) = \prod_{i=1}^M  [q_i(\thB)]^{b_i} \,
[1-q_i(\thB)]^{1-b_i}. \label{e:lik}
\end{equation}

\section{Parameter estimation}

The estimation problem is formulated in the Bayesian framework. The goal
is to compute the posterior PDF: the probability distribution of the
parameter vector $\thB$  conditional on the measurement vector
$\bb$. The posterior PDF provides a complete probabilistic
description of the information contained in the measurements about
the parameter vector $\thB$. The basic elements required to compute the
posterior distribution of are: (i) the prior distribution for the
parameter vector $\pi(\thB)$ and (ii) the likelihood function
$p(\bb|\thB)$. Given these quantities, Bayes rule can be used to
find the posterior PDF  as
\begin{equation}
p(\thB|\bb) = \frac{p(\bb|\thB)\pi(\thB)}{\int p(\bb|\thB)\pi(\thB)
d\thB}.
\end{equation}
The prior distribution $\pi(\thB)$ is typically non-Gaussian. For
example, the source position can be restricted to polygon regions,
while $Q_0$ and $V$ are strictly positive random variables.
Quantities of interest related to $\thB$ (e.g., the posterior mean,
variance) can be computed from the posterior PDF.

\begin{figure*}[t!]
\centering
\begin{tabular}{cc}
\includegraphics[height=7.0cm]{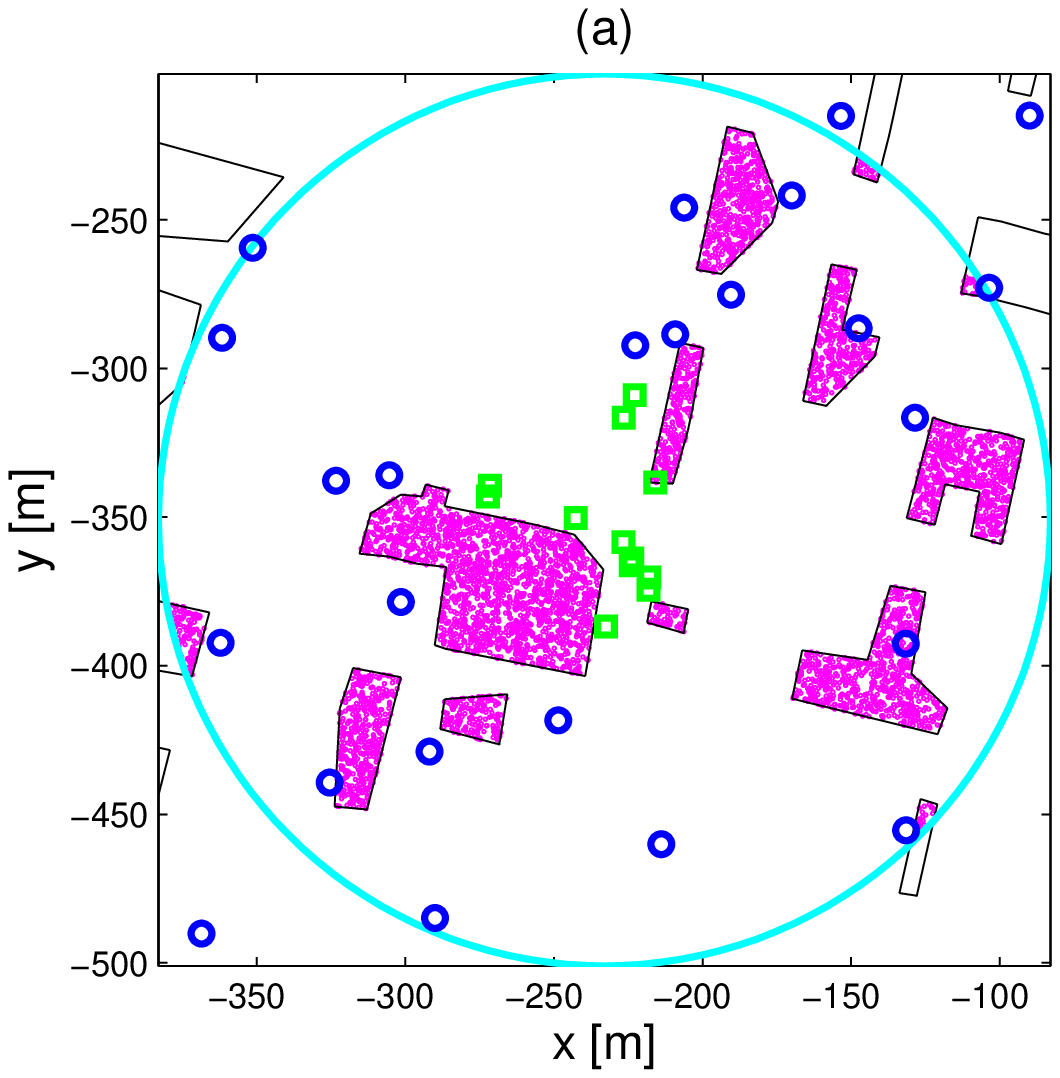} &
\includegraphics[height=7.0cm]{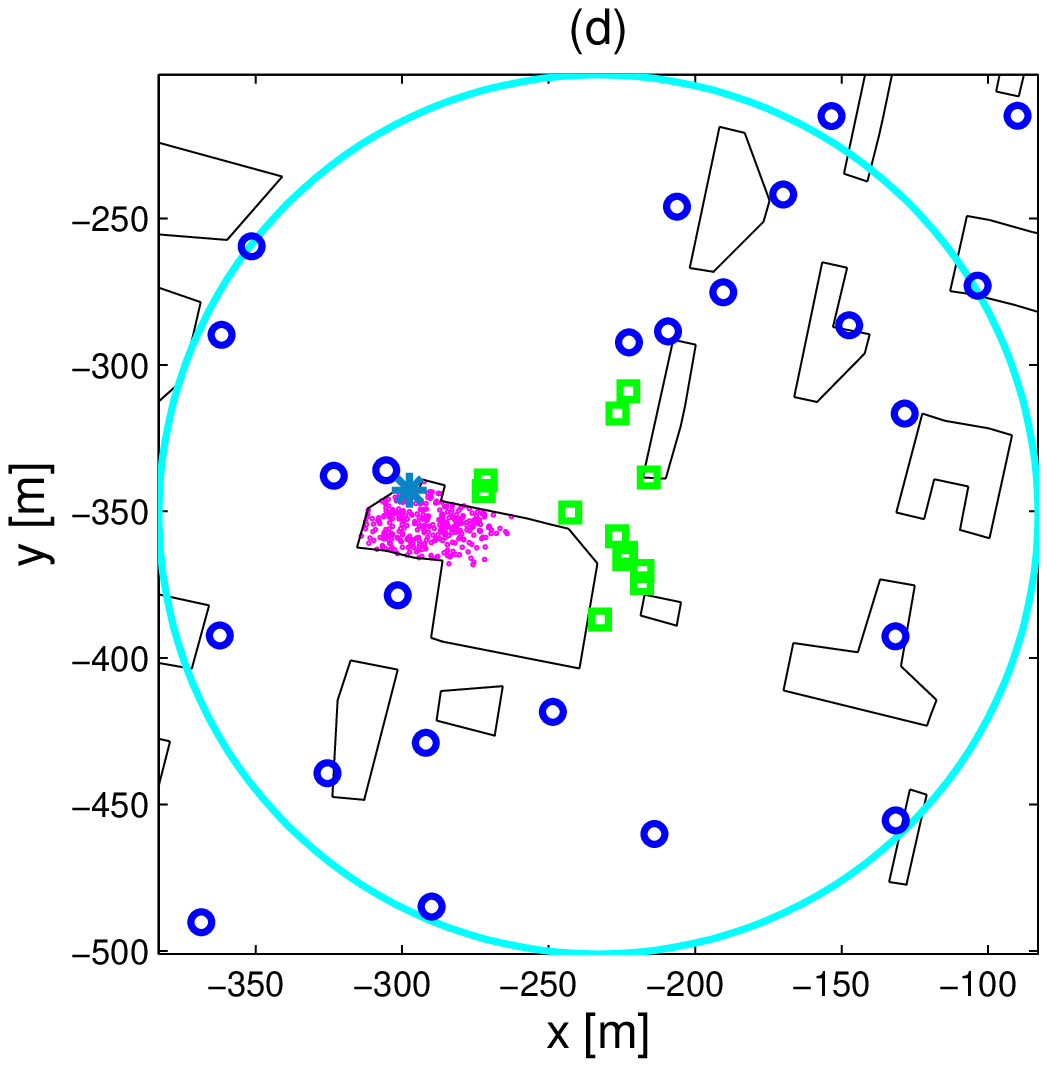}\\
\includegraphics[height=5.3cm]{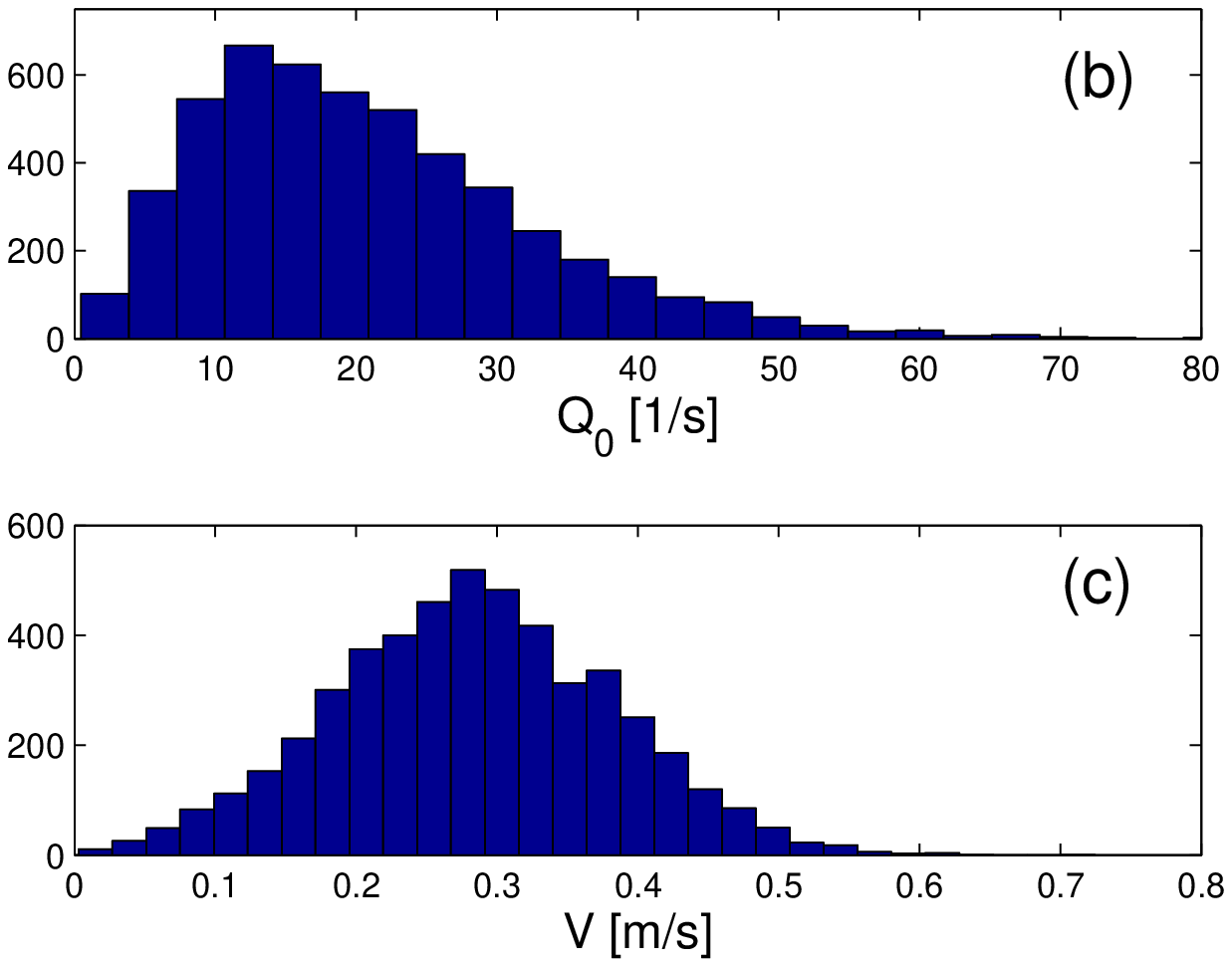} &
\includegraphics[height=5.3cm]{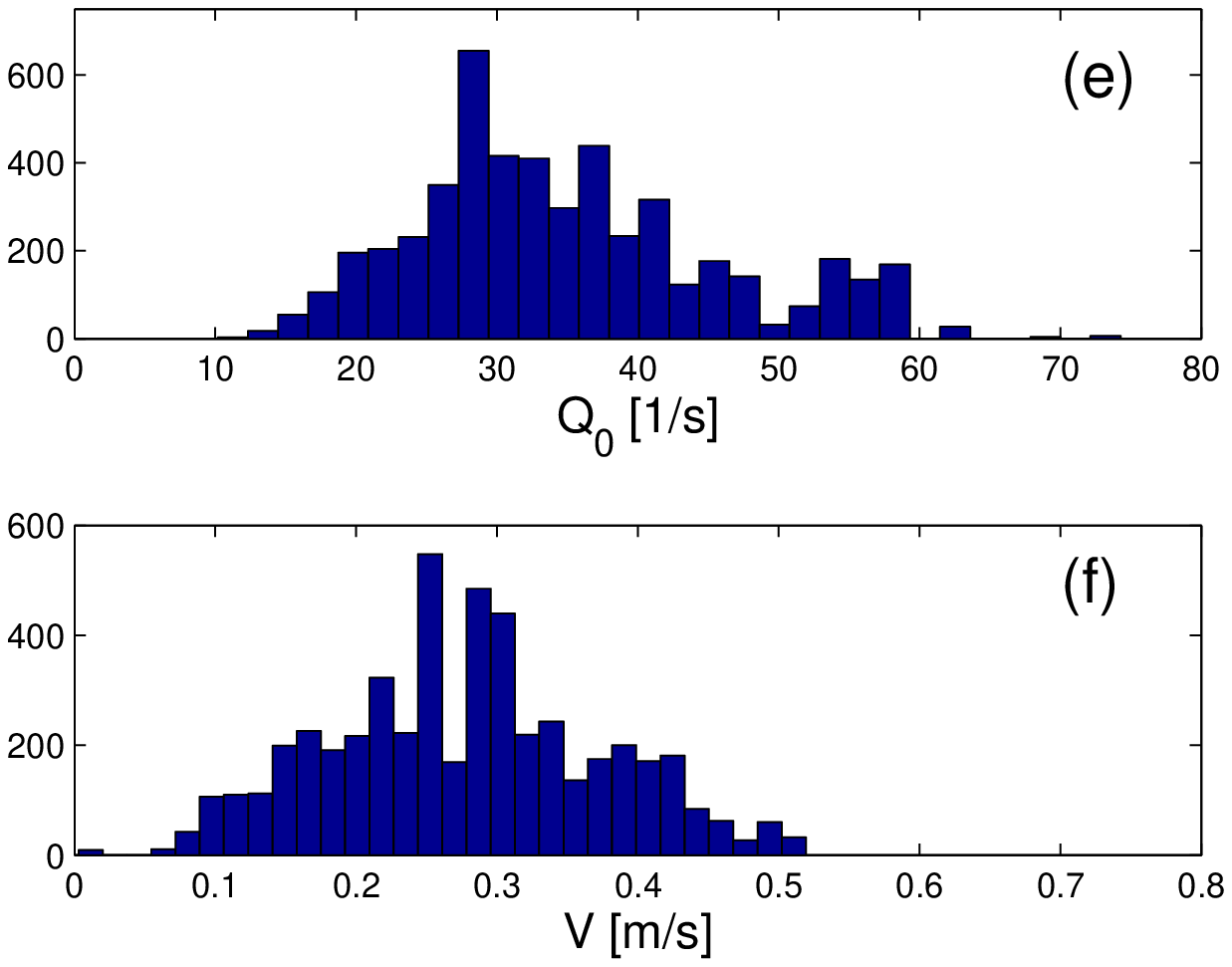}
\end{tabular}
 \caption{\footnotesize Estimation results for dataset 1: the left column is the prior PDF; the right column is the posterior
 PDF. Marginalised prior PDF $\pi(x_0,y_0)$ in (a) and the marginalised posterior PDF $p(x_0,y_0|\bb)$ in
 (d), approximated by random samples (indicated by scattered red dots).
Sensor locations (green squares are positive readings,
 blue circles are non-detections), as well as the building contours (black lines), also indicated in (a) and (d).
 Wind direction coincides with the $x$ axis. True source marked in (d) at $(-298.4,-342.6)$ (grey asterix). Figures (b) and (e) show the histograms corresponding to
 $\pi(Q_0)$ and $p(Q_0|\bb)$, respectively. Figures (c) and (f) show the histograms corresponding to $\pi(V)$ and $p(V|\bb)$, respectively.}
 \label{f:1}
\end{figure*}

Optimal Bayesian estimation is generally impossible because the
posterior PDF cannot be found in closed-form; this is certainly the
case for the likelihood function specified in Sec.\ref{s:2} and a
non-Gaussian prior $\pi(\thB)$. Hence we apply a Monte Carlo
approximation technique of the optimal Bayesian estimation, known as
importance sampling \cite{robert_casella}. This technique
approximates the posterior PDF by a weighted random sample
$\{w_n,\thB_n\}_{1\leq n \leq N}$, which is created as follows.
First, a sample $\{\thB_n\}_{1\leq n \leq N}$ is drawn from an
importance distribution $\varrho$, i.e., $\thB_n\sim \varrho(\thB)$,
for $n=1,\dots,N$. The unnormalised  weights are then computed as:
\begin{equation}
\tilde{w}_n = \frac{\pi(\thB_n)}{\varrho(\thB_n)}\,p(\bb|\thB_n)
\end{equation}
for $n=1,\dots,N$. Finally, the weights are normalised, i.e., $w_n =
\tilde{w}_n /\sum_{n=1}^N \tilde{w}_n$, for $n=1,\dots,N$. The
choice of importance distribution $\varrho$ plays a significant role
in the convergence of point estimators based on the approximation
$\{\thB_n\}_{1\leq n \leq N}$. Ideally $\varrho$ should resemble the
posterior. Since the posterior is unknown, good importance
distributions are often designed iteratively (population Monte Carlo
\cite{robert_casella},  progressive correction \cite{musso_et_al_00,
morelande_ristic_rad}). As $N\rightarrow \infty$, however, the
choice of $\varrho$ is less relevant and even the prior $\pi$ may be
used as an (admittedly inefficient) importance distribution; this is
done in our implementation for convenience. Furthermore, once the weighted random
sample $\{w_n,\thB_n\}_{1\leq n \leq N}$ is computed, resampling
(with replacement) is carried out \cite{pfbook} resulting in a
sample with uniform weights. This last step was carried out mainly
to improve the effect of visualisation of the posterior PDF (see
figures in the next section). In our implementation, the prior PDF
is adopted as:
\begin{equation}
\pi(\thB) = \pi(x_0,y_0)\, \pi(Q_0)\, \pi(V)
\end{equation}
where: $\pi(x_0,y_0)$ is a uniform distribution over designated
polygon areas (e.g., buildings); $\pi(Q_0)$ is a Gamma distribution
with shape $k$ and scale parameter $\eta$; $\pi(V)$ is a normal
distribution with mean $\bar{V}$ and standard deviation $\sigma_V$.

%The likelihood function in (\ref{e:lik}), as a probability density
%function, represents a precise probabilistic characterisation  of
%measurements. In practice, such precise (categorical)
%characterisations can cause problems in estimation, because any
%model deviation will result in distorted posterior PDFs. When
%dealing with experimental data, mathematical models such as
%(\ref{e:disp}) and (\ref{e:likf}), should be taken with

\section{Experimental results}

Three experimental datasets collected using a single moving binary sensor with unknown detection threshold were used in the paper to demonstrate the algorithm performance(further details of the experiments cannot be revealed on security grounds).  In all cases, algorithm parameters were
adopted as follows: $a=0.2$ m, $D=1$ m$^2$/s, $\tau = 1000$ s,
$t_0=1$ s,  sample size $N=5000$, shape parameter $k=3$, scale
parameter $\eta=7$, standard deviation of wind speed $\sigma_V =
0.2$ m/s. The wind conditions were different for the three datasets.
The mean wind direction was specified by angle $\alpha$, measured
anticlockwise from the $x$ axis. The values of wind parameters
$(\bar{V},\alpha)$ were $(0.28 \text{ m/s}, 195^\circ)$, $(0.12
\text{m/s},-10^\circ)$ and $(0.14 \text{m/s},150^\circ)$ for the
first, second and the third dataset, respectively.

Fig.\ref{f:1} shows the results obtained using dataset 1: the left
column displays the prior PDF $\pi(\thB)$; the right column - the
posterior PDF $p(\thB|\bb)$. Figs. \ref{f:1}.(a) and \ref{f:1}.(d)
display the top down view of the area where the experiment was
carried out. The placement and the readings of the binary sensor
are also marked (the number of sensor locations in dataset 1 is $M=45$).
The locations where the sensor reported positive readings (i.e., $b_i=1$), are marked by
green squares; the remaining (non-detecting) sensor locations are marked by blue circles. Based on prior knowledge (e.g., intelligence reports), figures (a) and (d) also indicate the circular area where the source must be located (circle drawn in cyan colour). The center of this circle is the mean position of the sensor locations with
positive readings; its radius is $150$m. Furthermore, figures (a)
and (d) also display the countours of the buildings (black lines).
Both the marginal prior $\pi(x_0,y_0)$ and the marginal posterior
PDF $p(x_0,y_0|\bb)$, are approximated by random samples marked by
scattered red dots. Assuming the source must be inside one of the
buildings, the marginal prior PDF $\pi(x_0,y_0)$ shown in figure (a)
is a uniform PDF over the intersection of the cyan color bounded
circular area and the area covered by the buildings. The marginal
posterior PDF $p(x_0,y_0|\bb)$, shown in figure (d), concentrates in
the upper left corner of one of the buildings, thus 
dramatically reducing the initial uncertainty in the source location. The
true source location is marked by an asterisk at $(-298.4,-342.6)$.
Figs.\ref{f:1}. (b) and (c) show the histograms of random samples
approximating $\pi(Q_0)$; and $\pi(V)$, respectively. Figs.
\ref{f:1}.(e) and \ref{f:1}.(f) display the histograms of random
samples approximating the marginal posteriors $p(Q_0|\bb)$ and
$\pi(V|\bb)$, respectively.

\begin{figure}[h!]
\centerline{\includegraphics[height=7.0cm]{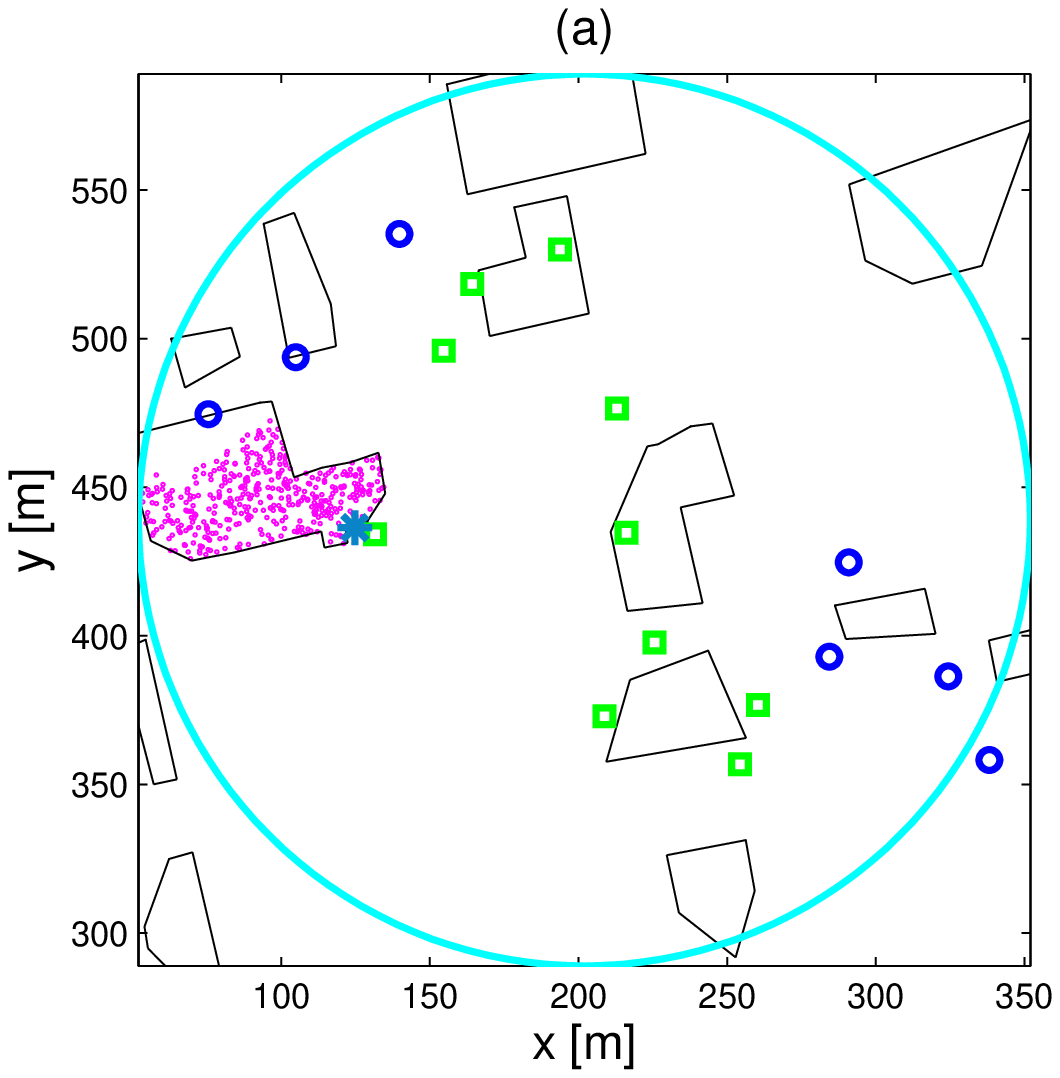}}
\centerline{\includegraphics[height=5.3cm]{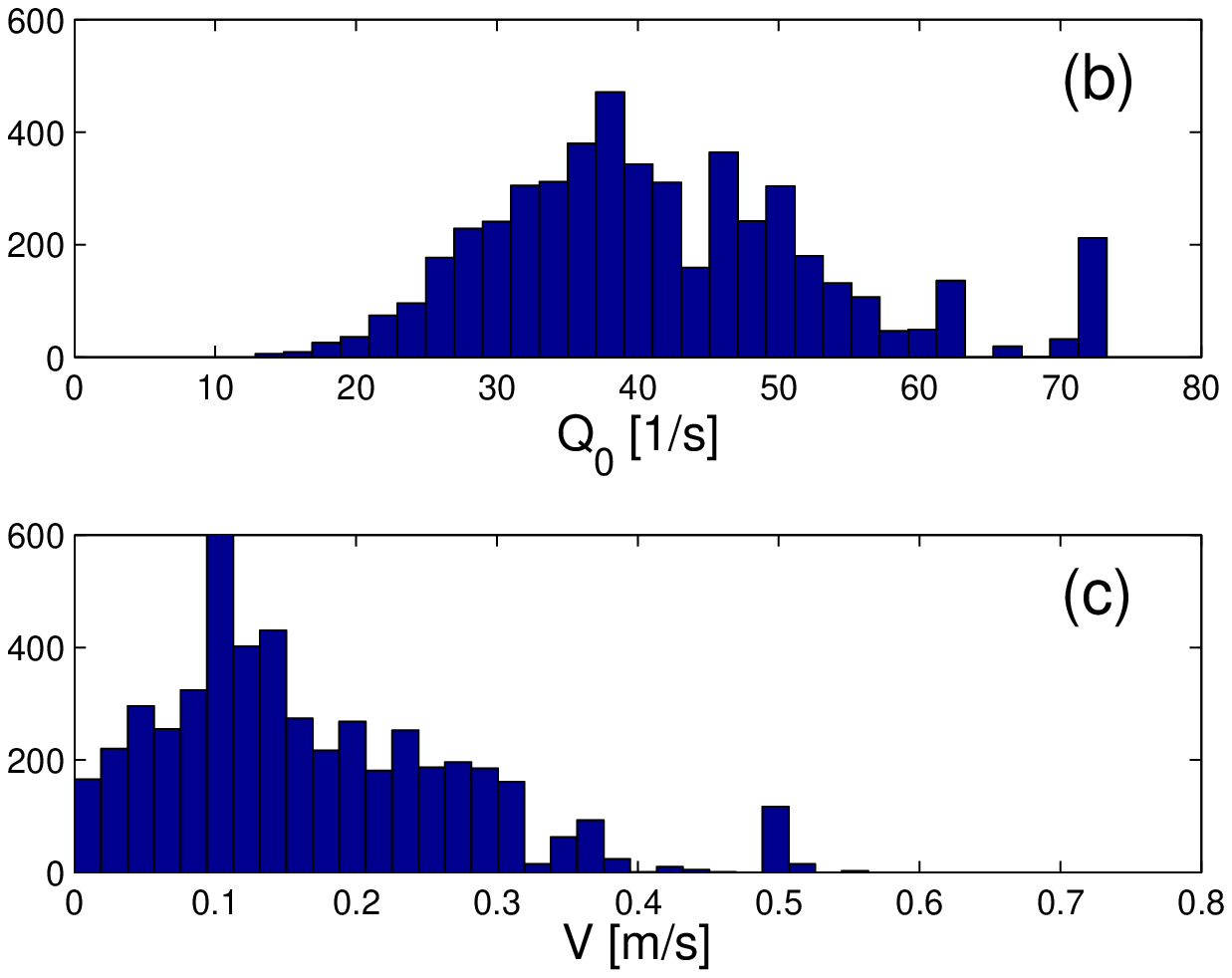}}
 \caption{\footnotesize Estimation results obtained by processing dataset 2: (a)  Marginalised posterior PDF $p(x_0,y_0|\bb)$
  (scatter plot, red dots); sensor locations (green squares are positive readings,
 blue circles are non-detections); building contours (black lines);
 true source location at $(125.6,436.6)$ (grey asterix).
 Wind direction coincides with the $x$ axis. Figures (b) and (c) display the histograms of random samples approximating $p(Q_0|\bb)$ and
  $p(V|\bb)$, respectively. }
 \label{f:2}
\end{figure}
\begin{figure}[h!]
\centerline{\includegraphics[height=7.0cm]{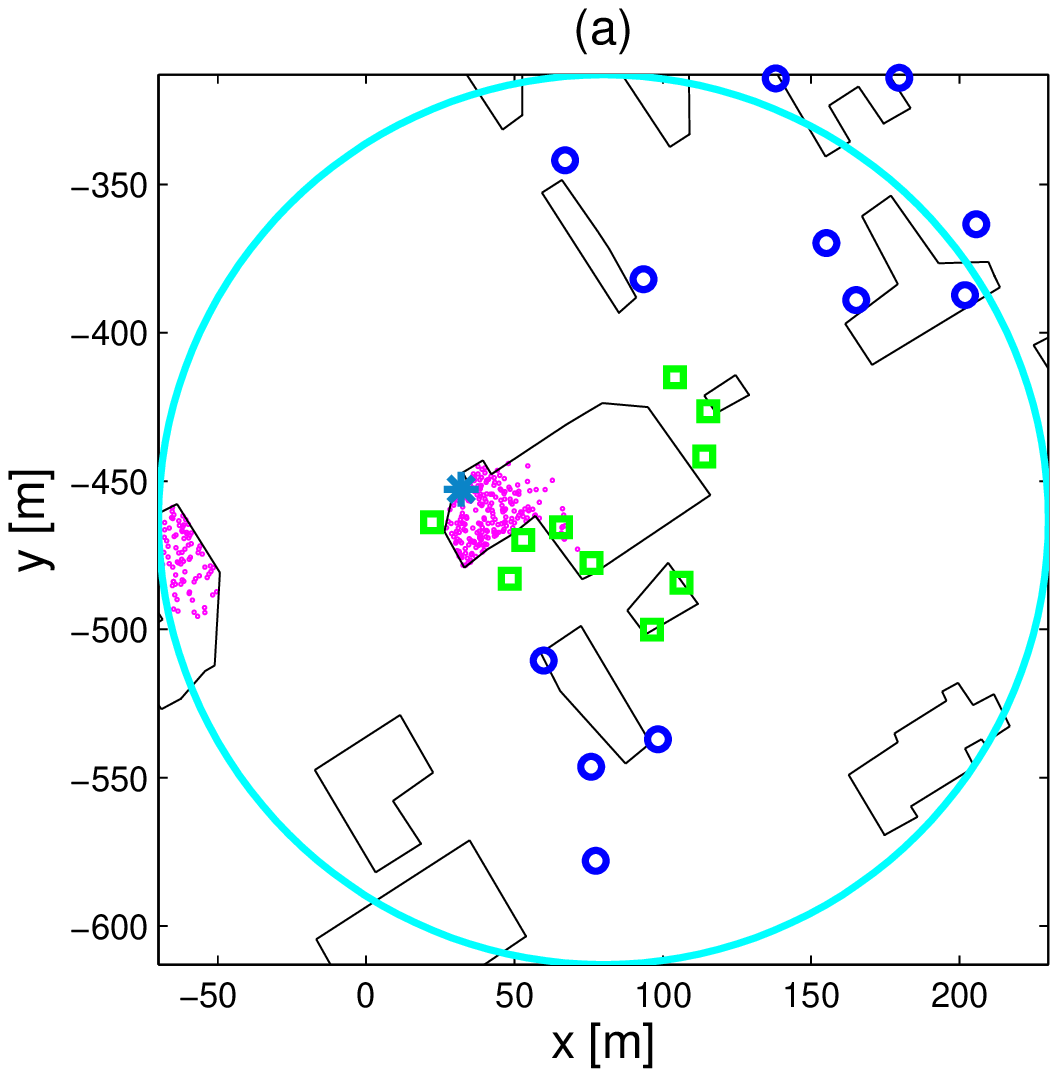}}
\centerline{\includegraphics[height=5.3cm]{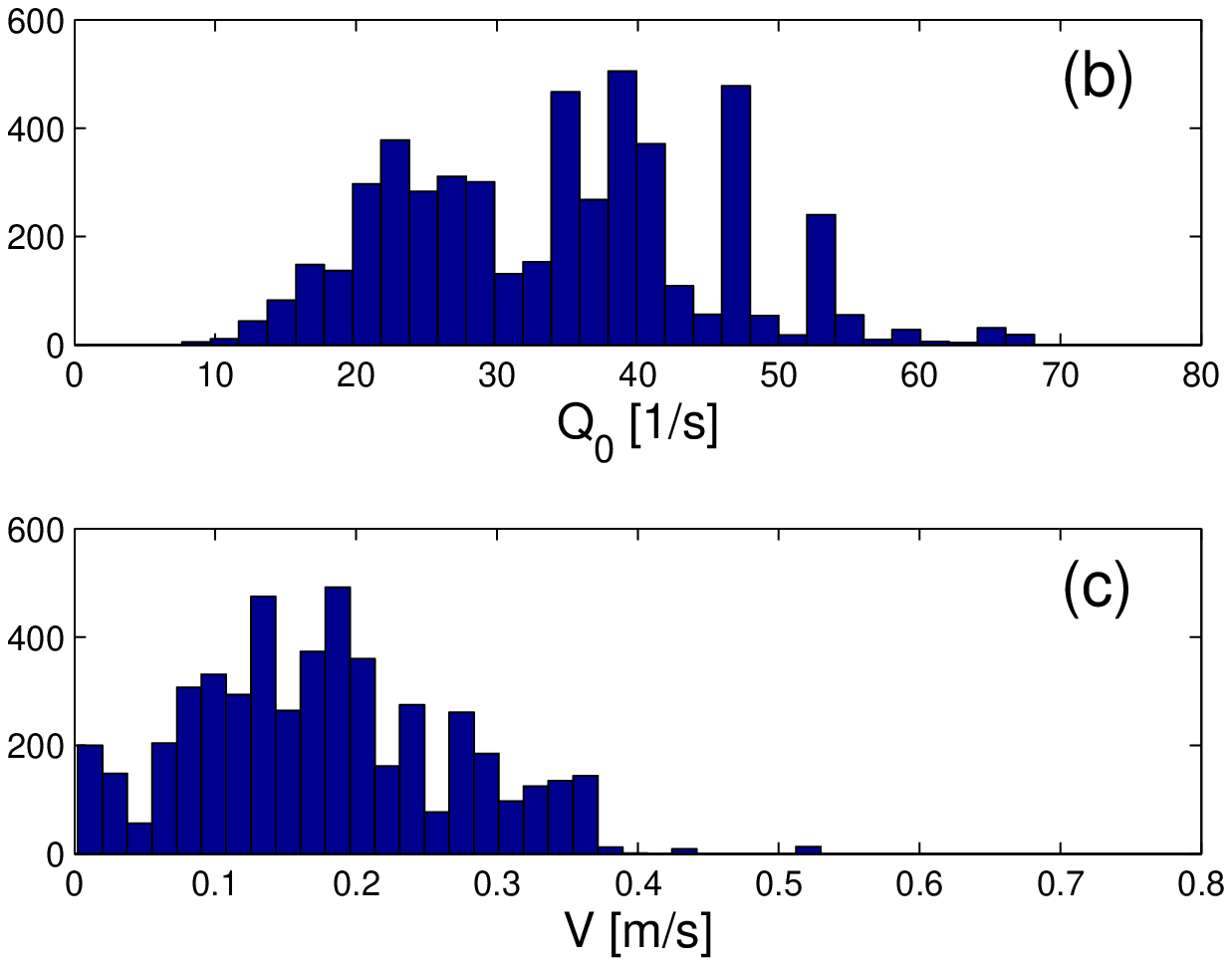}}
 \caption{\footnotesize Estimation results obtained by processing dataset 3: (a)  Marginalised posterior PDF $p(x_0,y_0|\bb)$
  (scatter plot, red dots); sensor locations (green squares are positive readings,
 blue circles are non-detections); building contours (black lines).
 True source location at $(31.2,-453.2)$ (grey asterix).
 Wind direction coincides with the $x$ axis.  }
 \label{f:3}
\end{figure}

The marginal posterior PDFs, obtained by processing dataset 2, are
shown in Fig.\ref{f:2}. The number of  sensor measurement locations in this case
was $M=25$. The initial random sample $\{\thB_n\}_{1\leq n \leq N}$
was created in the same manner as for the case of dataset 1. From
Fig.\ref{f:2}.(a) we can observe that the posterior PDF
$p(x_0,y_0|\bb)$ indicates fairly accurately the true source
location marked by an asterisk at coordinates $(125.6,436.6)$.

Finally, the marginal posterior PDFs, obtained by processing dataset
3, are shown in Fig.\ref{f:3}. The number of  sensor locations in
this case was $M=27$. The initial random sample $\{\thB_n\}_{1\leq n
\leq N}$ was created in the same manner as for the case of dataset
1. From Fig.\ref{f:3}.(a) we can observe that the support of the
marginal posterior PDF $p(x_0,y_0|\bb)$ indeed contains the true
source location at coordinates $(31.2,-453.2)$. However, the
conditions were such (the placement of sensor, the wind speed) that
some ambiguity in the source location remains.  The resulting
posterior PDF is bi-modal (two buildings contain scattered red
dots), suggesting that the source must be located in one of them.

\section{Conclusions}
\label{s:6}

The paper proposed a simple Bayesian estimation algorithm for
localisation of a continuous source of biochemical agent dispersing in the
atmosphere, using measurements collected at multiple locations by a single moving binary sensor whose detection threshold is unknown. The algorithm would also be applicable to  a single snapshot of the measurements from a network of identical binary sensors. The sensor detection threshold may be unknown because it may have been drifted due to temperature, humidity, ageing, etc. Another possible scenario where the detection threshold of a binary sensor  may be unknown is when a human, rather than a device, detects an odour at some locations but not  others. In this scenario, the person can easily make the binary measurements of ``detection'' or ``non-detection'' without knowing the exact detection threshold in terms of ppm or g/m$^3$ of the detected material. To enable source localisation in such scenarios, in our algorithm, we treat the source release rate, as well as the binary sensor threshold, as being unknown. Under these conditions, the algorithm can not estimate the absolute value of the source release rate and is only able to estimate the release rate normalised by the unknown sensor threshold. However, the algorithm can correctly estimate the location of the biochemical source.  The performance of the
algorithm is demonstrated using three experimental datasets
collected in a semi-urban environment. In all three cases, the
posterior density function included the true source location,
thereby validating the proposed algorithm. Future work will consider
introducing the uncertainty in the mean wind direction and more
detailed dispersion models for urban environments.

\section*{References}
%% `Elsevier LaTeX' style
\bibliographystyle{elsarticle-num}
\bibliography{binary_GPS}
%\bibliography{}

\end{document}